\documentclass[iop]{emulateapj-rtx4}
\usepackage{natbib}
\usepackage{color}
\usepackage{graphicx}
\newcommand{\Ka}{K$\alpha$}

\newcommand{\kTe}{$kT_{\rm e}$}
\newcommand{\kTinit}{$kT_{\rm init}$}
\newcommand{\net}{$n_{\rm e}t$}

\shorttitle{Thermal Emission from W49B with {\it NuSTAR}}
\shortauthors{Yamaguchi et al.}

\begin{document}

\title{Evidence for Rapid Adiabatic Cooling as an Origin of the Recombining Plasma in \\
the Supernova Remnant W49B Revealed by {\it NuSTAR} Observations}

\author{
Hiroya Yamaguchi\altaffilmark{1},
Takaaki Tanaka\altaffilmark{2}, 
Daniel R.\ Wik\altaffilmark{3}, 
Jeonghee Rho\altaffilmark{4,5},
Aya Bamba\altaffilmark{6,7},
Daniel Castro\altaffilmark{8}, \\
Randall K.\ Smith\altaffilmark{8},
Adam R.\ Foster\altaffilmark{8},
Hiroyuki Uchida\altaffilmark{2},
Robert Petre\altaffilmark{9},
Brian J.\ Williams\altaffilmark{9}
}
\email{yamaguchi@astro.isas.jaxa.jp}

\altaffiltext{1}{Institute of Space and Astronautical Science, JAXA, 3-1-1 Yoshinodai, Sagamihara, 
	Kanagawa 229-8510, Japan}
\altaffiltext{2}{Department of Physics, Kyoto University, Kitashirakawa-oiwake-cho, Sakyo-ku, 
	Kyoto 606-8502, Japan}
\altaffiltext{3}{Department of Physics \& Astronomy, University of Utah, 
	115 S.\ 1440 E.\ Salt Lake City, UT 84112}
\altaffiltext{4}{SETI Institute, 189 Bernardo Ave, Mountain View, CA 94043}
\altaffiltext{5}{SOFIA Science Center, NASA Ames Research Center, MS 232, Moffett Field, CA 94035}
\altaffiltext{6}{Department of Physics, The University of Tokyo, Bunkyo, Tokyo, 113-0033, Japan}
\altaffiltext{7}{Research Center for the Early Universe, School of Science, 
The University of Tokyo, 7-3-1 Hongo, Bunkyo-ku, Tokyo 113-0033, Japan}
\altaffiltext{8}{Harvard-Smithsonian Center for Astrophysics, 60 Garden Street, 
	Cambridge, MA 02138, USA}
\altaffiltext{9}{NASA Goddard Space Flight Center, Code 662, Greenbelt, MD 20771, USA}

\begin{abstract}

X-ray observations of supernova remnants (SNRs) in the last decade have shown that 
the presence of recombining plasmas is somewhat common in a certain type of object. 
The SNR W49B is the youngest, hottest, and most highly ionized among such objects, 
and hence provides crucial information about how the recombination phase is reached 
during the early evolutionary phase of SNRs.  
In particular, spectral properties of radiative recombination continuum (RRC) from Fe 
are the key for constraining the detailed plasma conditions. Here we present imaging and 
spectral studies of W49B with {\it Nuclear Spectroscopic Telescope Array (NuSTAR)}, 
utilizing the highest-ever sensitivity to the Fe RRC at $\gtrsim$\,8.8\,keV. 
We confirm that the Fe RRC is the most prominent at the western part of the SNR 
because of the lowest electron temperature ($\sim$\,1.2\,keV) achieved there. 
Our spatially resolved spectral analysis reveals a positive correlation between the electron 
temperature and the recombination timescale with a uniform initial temperature of $\sim$\,4\,keV, 
which is consistent with the rapid adiabatic cooling scenario as an origin of the overionization. 
This work demonstrates {\it NuSTAR}'s suitability for studies of thermal emission, 
in addition to hard nonthermal X-rays, from young and middle-aged SNRs.

\end{abstract}


%%%%%%%%%%%%
%%% Section 1 %%%
%%%%%%%%%%%%

\section{Introduction}
\label{sec:intro}

X-ray spectroscopy of supernova remnants (SNRs) allows us to investigate the thermal properties 
of the shocked gas (both swept-up ambient medium and supernova ejecta), providing a powerful 
probe for the SNR's environment and evolution history. In the last decade, studies with modern 
X-ray observatories have shown that a number of middle-aged SNRs contain recombining 
(overionized) plasmas, where the heavy element ions have been stripped of more electrons 
than they would be if the plasma is in ionization equilibrium 
\citep[e.g., IC\,443, W28, W44, N49: ][]{Yamaguchi09,Sawada12,Uchida12,Uchida15}. 
This fact implies that the presence of the `recombination phase' during the remnant evolution 
is somewhat common among a certain type of SNRs. Yet, detailed mechanisms that lead to 
the observed plasma properties are still poorly understood.

%%% Figure 1 %%%
\begin{figure*}[t!]
  \begin{center}
        \includegraphics[width=17.4cm]{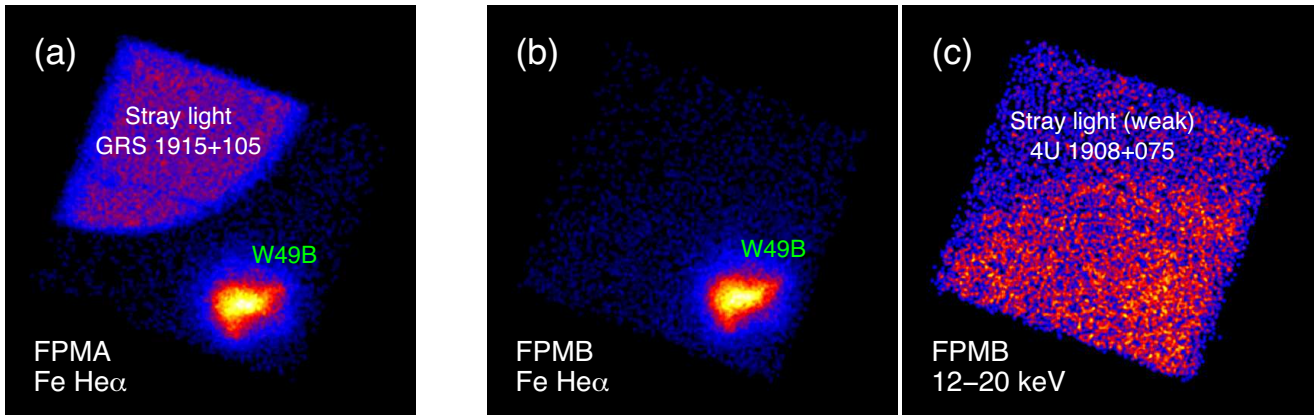}
        \vspace{2mm}
\caption{\footnotesize
Raw photon count images of (a) the FPMA in 6.4--6.8\,keV, (b) FPMB in 6.4--6.8\,keV, 
and (c) FPMB in 12--20\,keV, in the square root scale. 
The stray light features from GRS 1915+105 and 
4U 1908+075 are detected in the FPMA and FPMB, respectively. 
\label{image1}}
        \vspace{2mm}
  \end{center}
\end{figure*}

The SNR W49B is an extremely intriguing object in both thermal and nonthermal aspects, 
given that it is the most luminous in Fe K-shell emission \citep{Yamaguchi14b} 
and GeV $\gamma$-rays \citep{Abdo10,HESS18} among Galactic SNRs. 
This Letter focuses exclusively on its thermal aspect, 
whereas \cite{Tanaka18} present our new results on the nonthermal phenomena. 
Notably, W49B is the youngest \citep[1,000--6,000\,yrs: e.g.,][]{Pye84,Smith85,Zhou18} among 
the known SNRs in the recombining state, making its plasma extraordinarily hot and highly-ionized.
Previous {\it Suzaku} observations detected a strong radiative recombination continuum (RRC) 
of He-like Fe at $E_{\rm edge}$ $\approx$ 8.83\,keV, a key spectral feature to 
constraining the ionization state and electron temperature \citep{Ozawa09b}.  
There have been several attempts to determine the spatial distribution of the recombining 
plasma in W49B using {\it XMM-Newton} \citep{Miceli10} and {\it Chandra} \citep{Lopez13b,Zhou18}. 
Both observations indicated that the degree of overionization is more significant in the west than 
in the east. However, none of these works constrained the detailed plasma properties based on 
the realistic spectral modeling using the Fe RRC, because of the low signal-to-noise ratios 
near and above $E_{\rm edge}$.

Here, we present deep observations of W49B with {\it Nuclear Spectroscopic Telescope Array (NuSTAR)} 
that was recently performed with the aim of revealing the early evolutionary characteristics of 
the overionized SNRs. Although {\it NuSTAR} is designed to achieve a good sensitivity and 
an imaging capability to nonthermal radiation components in the hard ($\gtrsim$\,10\,keV) 
X-ray band (including both nonthermal continuum and radioactive decay lines of $^{44}$Ti), 
the `turnover' of the effective area in comparison to other X-ray telescopes takes place at 
$\sim$\,6.5\,keV \citep{Harrison13}, well below $E_{\rm edge}$ of the Fe RRC. 
Moreover, its angular resolution or half power diameter (HPD $\approx$ $1'$) is twice better than 
that of {\it Suzaku}, and reasonably smaller than the SNR's angular size ($4'$ $\times$ $3'$). 
Utilizing these capabilities, we perform the first spatially-resolved spectroscopy of W49B including 
the Fe RRC component.

This Letter is organized as follows.
In \S2, we describe details of our {\it NuSTAR} observations and data reduction. 
The screened data are analyzed and results are discussed in \S3.
Finally, we conclude this study in \S4. 
The errors quoted in the text and table and error bars given in the figures 
all represent a 1$\sigma$ confidence level.

%%%%%%%%%%%%
%%% Section 2 %%%
%%%%%%%%%%%%

\section{Observation and Data Reduction}
\label{sec:observation}

W49B was observed by {\it NuSTAR} on March 17--20, 2018
(Obs.\ ID: 40301001002) during Cycle 3 of the Guest Observer Program. 
We reprocessed the data using the {\tt nupipeline} task in the NuSTARDAS v.1.8.0 
software package with the calibration database (CALDB) released on April 19, 2018. 
We also filtered out periods when the background is high, resulting in the effective exposure of 122\,ks.  
The strictness of our filtering criteria is comparable to that of the {\tt saamode\,=\,optimized} and  
{\tt tentacle\,= yes} options in the {\tt nupipeline} routine.

Figure\,1a shows a photon count image of the Focal Plane Module A (FPMA) 
in the 6.4--6.8\,keV band, corresponding to the energies of the Fe He$\alpha$ lines. 
W49B is observed using the Det\,0 chip (bottom right in the figure) where the optical axis is located. 
The strong stray light from the black hole binary GRS 1915+105 is detected at 
the off-source regions, which does not affect the analysis of W49B.  
Figures~1b and 1c are FPMB images in 6.4--6.8\,keV and 12--20\,keV, respectively. 
Unlike the FPMA, the on-axis Det\,0 chip suffers from the stray light from 
the high mass X-ray binary 4U 1908+075. However, its flux level is relatively low, 
so the feature is visible only in the hard X-ray band. We appropriately take into account 
this stray light effect in the subsequent background estimate and spectral analysis.

%%%%%%%%%%%%
%%% Section 3 %%%
%%%%%%%%%%%%

\section{Results and Discussion}
\label{sec:results}

%%% Figure 2 %%%
\begin{figure}[t!]
  \begin{center}
          %\vspace{2mm}
        \includegraphics[width=7.6cm]{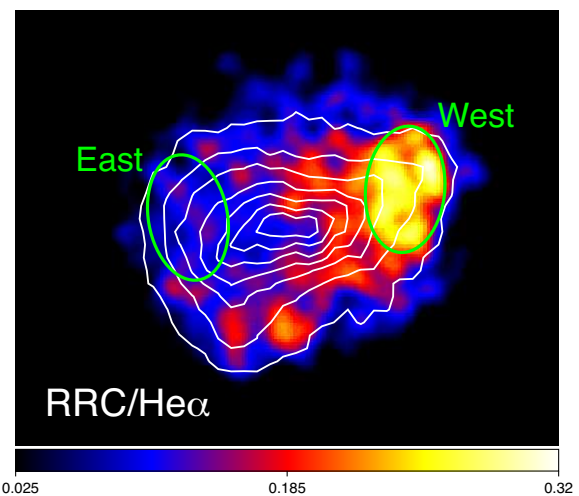}
\caption{\footnotesize
Spatial distribution of the flux ratio between the Fe RRC (8.8--10\,keV) and 
Fe He$\alpha$ (6.4--6.8\,keV) given in the linear scale.  
The ratios from the FPMA and FPMB are averaged and then smoothed. 
The overplotted contours are the exposure corrected Fe He$\alpha$ flux image. 
The green ellipses indicate where the spectra shown in Figure\,3 are extracted. 
\label{image2}}
\vspace{-0.4cm}
  \end{center}
\end{figure}

In Figure\,2, we show an image of the Fe RRC-to-He$\alpha$ flux ratio, 
which is generated by dividing the exposure corrected image in 8.8--10\,keV 
by that in 6.4--6.8\,keV and merging the data from the FPMA and FPMB. 
An enhancement of the Fe RRC (implying a large degree of overionization) is found at the west rim, 
consistent with the previous observations \citep{Miceli10}. For more quantitative study, 
we analyze spectra from two representative regions, East and West, labeled in Figure\,2. 
The background spectra, consisting of instrumental background, unresolved X-ray 
background, and stray light components, are generated using the {\tt nuskybgd} script, 
whose details are described in \cite{Wik14}.

%%% Figure 3 %%%
\begin{figure*}[t!]
  \begin{center}
        \includegraphics[width=17.2cm]{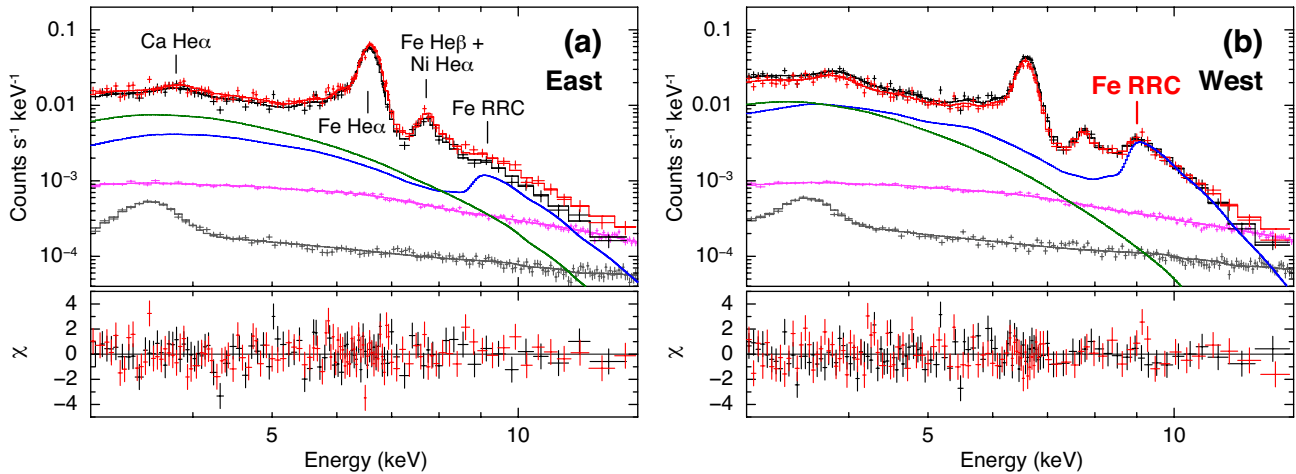}
\caption{\footnotesize
{\it NuSTAR} spectra of the East (a) and West (b), whose locations are given in Figure\,2. 
Black and red are the FPMA and FPMB, respectively. 
The lower panels show residuals from the best-fit models obtained using the C statistic method, 
where the background data generated with the {\tt nuskybgd} script
(gray and magenta for the FPMA and FPMB, respectively) are simultaneously fitted 
with the source data. The spectra are rebinned for clarity, although unbinned data are 
used in the actual analysis.
The green and blue curves indicate the contributions of the bremsstrahlung from H and He 
and the RRCs from heavy elements (C--Ni), respectively, to the FPMA spectra. 
\label{spectra}}
\vspace{-0.4cm}
  \end{center}
\end{figure*}

%%% Table 1 %%%
\begin{table*}
\begin{center}
\caption{Best-fit spectral parameters.
  \label{tab}}
  %\scriptsize
  %\tiny
  \begin{tabular}{lccccc}
\hline \hline
Parameter & \multicolumn{2}{c}{East} & & \multicolumn{2}{c}{West} \\
\cline{2-3} \cline{5-6}
& $\chi ^2$ & c-stat & & $\chi ^2$ & c-stat \\
\hline
\kTe\ (keV) & $1.84 \pm 0.05$ &  $1.86 \pm 0.05$ & & $1.20 \pm 0.04$ & $1.19 \pm 0.04$ \\
$kT_{\rm init}$ (keV) & $4.77 \pm 0.25$ & 4.77 (fixed)$^a$ & & $3.80_{-0.54}^{+0.80}$ & $3.73_{-0.49}^{+0.69}$ \\
$n_e t$ ($10^{11}$\,cm$^{-3}$\,s) & $6.32_{-0.34}^{+0.43}$ & $6.12_{-0.25}^{+0.32}$ & & $1.83_{-1.54}^{+1.65}$ & $1.66_{-1.45}^{+1.47}$  \\
Ca (solar)$^b$ & $3.9 \pm 0.8$ & $3.7 \pm 0.8$ & & $3.7 \pm 0.5$ & $3.5 \pm 0.5$ \\
Cr (solar)$^b$ & $12 \pm 3$ & $12 \pm 3$ & & $5.1_{-1.8}^{+2.0}$ & $5.8_{-1.8}^{+2.0}$ \\
Mn (solar)$^b$ & $69 \pm 8$ & $67 \pm 8$ & & $20 \pm 4$ & $21 \pm 4$ \\
Fe (solar)$^b$ & $6.4 \pm 0.4$ & $6.1 \pm 0.4$ & & $2.2 \pm 0.4$ & $2.2 \pm 0.4$ \\
Ni (solar)$^b$ & $13 \pm 3$ & $13 \pm 3$ & & $3.3_{-1.2}^{+1.4}$ & $3.4_{-1.2}^{+1.4}$ \\
Norm FPMA$^c$ & $8.09 \pm 0.33$ & $8.17 \pm 0.33$ & & $31.5 \pm 1.6$ &  $32.5 \pm 1.6$ \\
Norm FPMB$^c$ & $9.38 \pm 0.38$ & $9.56 \pm 0.39$ & & $30.9 \pm 1.6$ &  $32.2 \pm 1.6$ \\
Offset FPMA (ch)$^d$ &  --2.05 & --2.00 & & --1.23 & --1.15  \\
Offset FPMB (ch)$^d$ &  --1.48 & --1.45 & & --1.90 & --1.73  \\
\hline
$\chi ^2$ & 291 & 1058 & & 250 & 1009 \\
c-stat & --- & 1010 & & --- & 1064 \\
d.o.f. & 267 & 968 & & 260 & 969 \\
\hline
\end{tabular}
\tablecomments{
``$\chi ^2$'' \& ``c-stat'' in the second row indicate the statistical methods of chi-squared and C statistic, respectively. \\
$^a$Fixed to the best-fit value from the $\chi ^2$ method, because otherwise the value is not constrained at all. \\
$^b$Values relative to the solar abundances of \cite{Wilms00}. \\
$^c$Normalizations are independently fitted for the FPMA and FPMB. 
The unit is $10^{-17} / (4\pi D^2) \cdot \int n_{\rm e} n_{\rm H} dV$ (cm$^{-5}$), 
where $D$ is the distance to the source and $\int n_{\rm e} n_{\rm H} dV$ is the volume emission measure. \\
$^d$In {\it NuSTAR}, the width of each single pulse-height channel corresponds to the photon energy of 40\,eV.
}
\end{center}
\end{table*}

%%% Figure 4 %%%
\begin{figure*}[t!]
  \begin{center}
        \includegraphics[width=17.4cm]{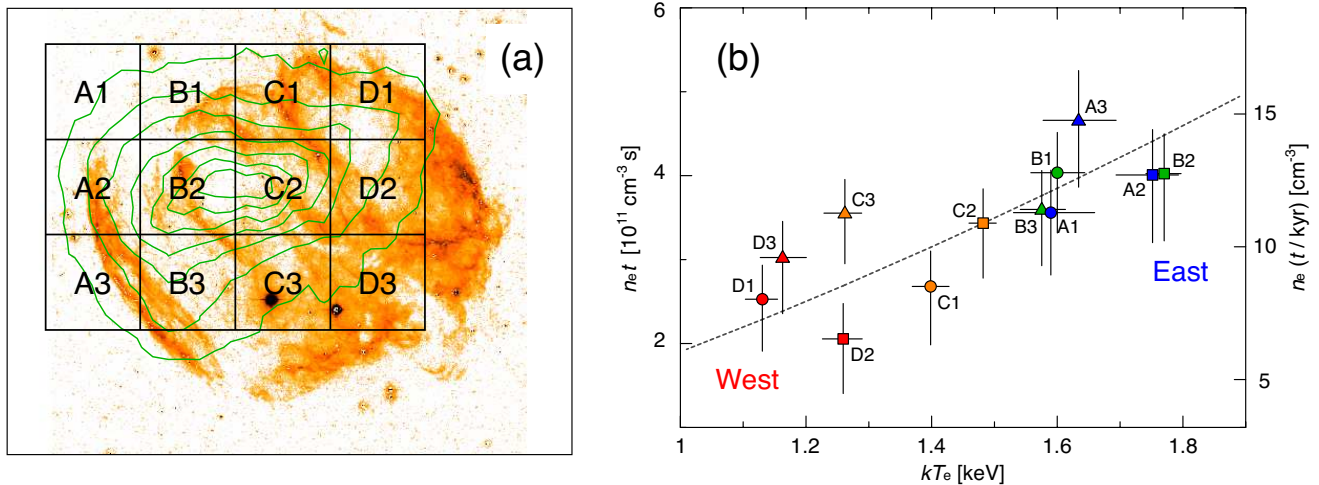}
\caption{\footnotesize
(a) The $1'$\,$\times$\,$1'$ box regions used for the spatially-resolved spectral analysis. 
The green contours are the exposure corrected Fe He$\alpha$ flux map 
(same as the white contours in Figure\,2). 
The color image is obtained with {\it Wide Field Infrared Camera (WIRC)} observations 
of the 1.64 $\mu$m [Fe\,{\footnotesize II}] band, where background stars are 
subtracted using the DAOPHOT method that is described in \cite{Rho01}.
\ 
(b) Relationship between the electron temperature (bottom) and the recombination timescale (left) 
or the corresponding electron density (right) obtained from the spatially-resolved spectral analysis. 
The region names labeled near the data points correspond to those given in panel\,(a). 
The dashed line is a power-law function with an index of 1.5, 
whose normalization is fitted to match the data.
\label{plot}}
  \end{center}
\end{figure*}

The spectra of the FPMA (black) and FPMB (red) from each source region are given in Figure\,3. 
The gray and magenta data points are the background spectra generated for the FPMA and FPMB, 
respectively, the latter showing the higher flux due to the stray light from 4U 1908+075.
The prominent RRC is confirmed in the West spectra (Figure\,3b), 
whose peak height is, surprisingly, comparable to that of the emission at $\sim$\,7.8\,keV 
(a mixture of Fe He$\beta$ and Ni He$\alpha$ lines). Moreover, the RRC has a steeper slope 
in the West, suggesting a lower electron temperature achieved there. 
To verify this, we fit the spectra with a recombining plasma model, 
{\tt vvrnei} %\footnote{https://heasarc.gsfc.nasa.gov/xanadu/xspec/manual/node144.html} 
in the {\tt XSPEC} package \citep{Arnaud96} based on the latest atomic database 
AtomDB version 3.0.9.$\!$\footnote{http://www.atomdb.org} 
The important parameters are the current electron temperature (\kTe), initial temperature (\kTinit) that 
determines the ionization balance before the abrupt plasma cooling, and recombination timescale (\net) 
that is the product of the electron density ($n_{\rm e}$) and the time elapsed after the abrupt cooling ($t$). 
Other free parameters are all listed in Table\,1. Abundances of unlisted heavy elements (e.g., Si, S)
are fixed to the solar values of \cite{Wilms00}. 
Normalizations of the FPMA and FPMB are fitted independently, following the instructions by  
{\it NuSTAR} Science Operations Center (SOC).$\!$\footnote{https://heasarc.gsfc.nasa.gov/docs/nustar/analysis/} 
A foreground absorption is accounted for using the {\tt tbabs} model \citep{Wilms00} 
with a fixed hydrogen column density of $5 \times 10^{22}$\,cm$^{-2}$ \citep{Keohane07}. 
We also try a larger value of $8 \times 10^{22}$\,cm$^{-2}$, which was reported in more recent 
work \citep{Lopez13a}, and find that this difference does not substantially affect our results.
Finally, we allow for an offset in the photon-energy to the pulse-hight relationship 
to account for possible gain calibration uncertainties.

The spectral fitting with the model described above is performed using two different statistical methods: 
(1) chi-squared ($\chi ^2$) statistic on background-subtracted, binned spectra, and 
(2) C statistic on background-unsubtracted, unbinned spectra \citep{Cash79}. 
For the latter, the background spectra are modeled by phenomenological functions 
and simultaneously fitted with the source data. The best-fit results are given in Table~1, 
which are obtained with reasonable goodness of fit ($\chi ^2$/d.o.f $<$ 1.1). 
In particular, the Fe RRC features are well reproduced in both regions. 
We find no significant difference between the results from the two statistical methods.

The low electron temperature and low recombination timescale 
(hence the substantial overionization) are confirmed at the West. 
At the observed temperatures, the intermediate-mass elements, such as Si and S, require 
\net\ $\gtrsim$ $10^{12}$\,cm$^{-3}$\,s to achieve an ionization equilibrium \citep{Smith10}. 
This indicates that these elements are also recombining in this SNR and thus 
the continuum emission in the {\it NuSTAR}'s energy band is dominated by their RRCs 
particularly at the West. %(i.e., not bremsstrahlung from H and He). 
This point is illustrated in Figure\,3, where the green and blue curves represent the contributions 
of the bremsstrahlung from H and He and the RRCs from heavy elements (C to Ni), respectively. 
The abundances given in Table~1 should, therefore, be regarded as relative values to the 
intermediate-mass elements (not to H), whose abundances are fixed to the solar values in our analysis. 
Another noteworthy fact is that the electron temperatures we have constrained are 
lower than some previous measurements \citep[2--3\,keV: e.g.,][]{Miceli06,Lopez09a}. 
This discrepancy can also be explained by the RRC effect; in these previous work, 
the spectra were modeled by ionization equilibrium plasmas, where the broadband 
continuum are dominated by the bremsstrahlung. 
Since the bremsstrahlung continuum has no spectral edge in the X-ray band, 
the attempt to reproduce the RRC-dominant spectrum with equilibrium plasma models
may have resulted in the overestimate of the electron temperature. 
On the other hand, our results are consistent with the electron temperatures of the `hot component' 
($kT_{\rm h}$) measured by \cite{Zhou18}, where recombining plasma models were used.

We obtain extremely high Mn abundances (Mn/Fe $\approx$ 10) from both regions, 
inconsistent with previous work \citep[e.g.,][]{Hwang00,Zhou18}. 
We suspect that this peculiar result is due to incomplete calibration of the line spread function, 
given that the Mn He$\alpha$ emission lies in the low energy tail of the stronger Fe He$\alpha$ lines. 
We do not investigate this problem in further details, because it does not affect the measurement of 
the other parameters, such as the electron temperature and recombination timescale, 
and the abundance structure is out of the scope of this Letter.

Finally, we perform spatially-resolved spectral analysis by dividing the Fe-rich regions into twelve 
$1'$\,$\times$\,$1'$ boxes (the size comparable to the {\it NuSTAR}'s HPD) indicated in Figure\,4a, 
where a 1.64-$\mu$m [Fe\,{\footnotesize II}] image from {\it Wide Field Infrared Camera (WIRC)} 
is also shown. The [Fe\,{\footnotesize II}] emission is a dominant cooling line from 
the interstellar medium with a density of 30--$10^4$\,cm$^{-3}$ and a temperature of 
$10^{3}$--$10^{5}$\,K \citep{Hewitt09}. 
Given that the best-fit \kTinit\ values for the East and West are comparable to each other ($\sim$\,4\,keV), 
we assume a common initial temperature shared over the entire SNR, and fit the 24 spectra 
(12 regions $\times$ 2 modules) simultaneously by linking \kTinit\ among the regions. 
The other parameters in Table~1 are all independently fitted. 
This analysis obtains \kTinit\ = $3.84_{-0.24}^{+0.18}$~keV with $\chi ^2$/d.o.f = 2745/2616. 
The values of \kTe\ and \net\ derived for each region are plotted in Figure\,4b, 
showing a clear correlation between the two quantities. 
In fact, we obtain a large positive correlation coefficient of 0.78.

The observed correlation suggests that lower electron temperatures are achieved in the lower density regions, 
qualitatively consistent with a rapid adiabatic expansion scenario as an origin of the overionization 
\citep{Itoh89,Yamaguchi09}. This scenario requires dense circumstellar matters (CSM) present 
close to the pre-explosion massive star, so that both CSM and supernova ejecta get shock heated 
and highly ionized shortly after the progenitor explosion \citep{Moriya12}. When the blast wave 
breaks out into the surrounding low-density region, the plasma cools rapidly \citep{Itoh89,Shimizu12}. 
Such density distribution can naturally be explained if the progenitor is a red supergiant (RSG), 
because the main-sequence wind can form a wind-blown cavity around the dense CSM of 
the RSG wind \citep[e.g.,][]{Dwarkadas05}. 
The massive progenitor scenario is also consistent with the presence of the dense wind-blown shell 
observed in the infrared image of the [Fe\,{\footnotesize II}] emission \citep{Keohane07}, 
although a Type Ia supernova origin was recently suggested for this remnant \citep{Zhou18}.

In adiabatic processes of an ideal monatomic gas in a closed system, $TV^{\gamma -1}$ is conserved, 
where $T$, $V$, and $\gamma$ (= 5/3) are the gas temperature, volume, and adiabatic index, respectively. 
Therefore, if the uniform temperature ($T_{\rm init}$) and density ($n_{\rm init}$) were achieved in 
the initially shock-heated materials, and if the rapid adiabatic cooling took place at the same time 
throughout the SNR, then the relationship $n_{\rm e}$ $\propto$ $T^{1.5}$ is expected 
among arbitrary fluid elements that have experienced the cooling. 
This expectation is confirmed in our spatially-resolved spectral analysis. 
A power-law function with an index of 1.5 (the dashed line in Figure\,4b) 
appropriately fits the observed relation between \kTe\ and $n_{\rm e}$ 
(which is derived simply from the recombination timescale), 
although, in reality, the SNR would have evolved through more complex paths than our assumptions. 
For instance, the initial condition ($T_{\rm init}$ and $n_{\rm init}$) might not have been uniform, 
and $T_{\rm e}$ and $n_{\rm e}$ might have changed even after the rapid cooling. 
Future theoretical work accounting for more realistic SNR evolution, 
such as those performed by \cite{Zhou11} and \cite{Slavin17}, would be helpful 
for detailed comparison between the observed and predicted plasma properties.

Recent studies of several other middle-aged SNRs suggest thermal conduction into the surrounding 
cold gas as a predominant origin of the overionization, given the fact that the recombining plasmas 
are localized near the molecular clouds \citep[e.g., G166.0+4.3, W28:][]{Matsumura17a,Okon18}. 
(See also \cite{Kawasaki05} for the thermal conduction scenario originally applied to W49B 
based on {\it ASCA} results.) 
In W49B, on the other hand, the electron temperature gradually goes down from the east 
to the west (Figure\,4b), and the plasma condition does not seem to be correlated with 
the ambient cold gas density that is represented by the infrared emission (Figure\,4a). 
This is another piece of evidence that adiabatic expansion is a more suitable explanation 
for the plasma overionization observed in this SNR. 
In fact, previous X-ray observations of W49B indicate a lower ambient density at the west than at the east, 
so the adiabatic cooling can take place more efficiently at the former \citep{Miceli10,Lopez13b}. 
The plausibility of this scenario is also confirmed by previous hydrodynamical simulations, 
where the density structure around this SNR is introduced as an initial condition \citep{Zhou11}.

%%%%%%%%%%%%
%%% Section 4 %%%
%%%%%%%%%%%%

\section{Conclusions}
\label{sec:conclusions}

We have presented the {\it NuSTAR} observations of W49B, focusing on its thermal aspect. 
A clear enhancement of the Fe RRC is observed at the western part of the SNR. 
Our spatially-resolved spectroscopy has revealed a positive correlation between 
the electron temperature and the recombination timescale (or the electron density), 
with a gradient from the west (lower) to the east (higher). 
The result can naturally be explained when the rapid adiabatic cooling is assumed to be 
a predominant origin of the overionization. The initial plasma 
temperature just before the rapid cooling took place (\kTinit) is estimated to be $\sim$\,4\,keV.
There is no spatial correlation between the plasma condition and the ambient cold gas distribution, 
making the thermal conduction scenario unlikely as a major driver of the rapid cooling in this SNR.

This work has newly expanded the capability of {\it NuSTAR}. 
Although this mission has so far focused on {\it nonthermal} phenomena in SNRs
\citep[e.g.,][]{Grefenstette15,Grefenstette17,Lopez15}, 
its large effective area and low background in 6--10\,keV are remarkably 
suitable for detecting {\it thermal} emission in this energy band 
(i.e., both lines and RRCs from the Fe-peak elements). 
This advantage can be utilized for observations of other SNRs, such as those emitting 
strong Ni \Ka\ lines \citep[e.g., Kepler, 3C\,397:][]{Park13,Yamaguchi15}, whose spatial 
distribution provides crucial information about the Type Ia supernova explosion mechanism. 
Searching for recombining plasmas in other SNRs \citep[e.g.,][]{Bamba18} would also be 
feasible for this observatory.

%%%%%%%%%%%%
% Acknowledgment %
%%%%%%%%%%%%

\acknowledgments

We are grateful to Karl Forster and Brian Grefenstette at {\it NuSTAR} SOC for 
their kind support for proposal preparation and observation planning as well as 
helpful advice on data screening and analysis. 
We also thank Keith Arnaud for his prompt response to our request to fix 
problems discovered in AtomDB-based plasma models in {\tt XSPEC}. 
%$\!$\footnote{https://heasarc.gsfc.nasa.gov/xanadu/xspec/issues/issues.html}
Useful information about {\it Suzaku} data analysis was provided by 
Shigeo Yamauchi, Masayoshi Nobukawa and Katsuji Koyama, 
when the observation proposal was prepared. 
We also acknowledge helpful discussion with Makoto Sawada. 
This work was supported by the NASA data analysis funding award for the {\it NuSTAR} Cycle-3 observing program, 
and made use of data from the {\it NuSTAR} mission, a project led by the California Institute of Technology, 
managed by the Jet Propulsion Laboratory, and funded by the National Aeronautics and Space Administration. 
This research has made use of the {\it NuSTAR} Data Analysis Software (NuSTARDAS) jointly developed by 
the ASI Science Data Center (ASDC, Italy) and the California Institute of Technology (USA).

\bigskip

%%% Refs %%%
%\bibliography{ms}

\end{document}